# Rejoinder to the Response to 'Comment on a recent conjectured solution of the three-dimensional Ising model'

Fa Yueh Wu[1], Barry M. McCoy[2], Michael E. Fisher[3] and Lincoln Chayes[4]

[1]Department of Physics, Northeastern University, Boston, Massachusetts 02115, USA
[2]Institute for Theoretical Physics, State University of New York, Stony Brook, New York 11794-3840, USA
[3]Institute for Physical Science and Technology, University of Maryland, College Park, Maryland 20742, USA
[4]Department of Mathematics, University of California, Los Angeles, California 90059-1555, USA

We add here a few sentences concerning the author's Response [1] to our Comment [2] criticizing his original claims regarding his conjectured solution of the three-dimensional Ising model [3].

First, we stand by our summary in [2] where the main purpose was to refute claims made in [3] on the basis of a putative 4-dimensional integral representation. In summarizing his rebuttal, Professor Zhang now admits that "more research" is needed.

He goes on, however, to assert that "the correct reproduction of the high-temperature expansion cannot be a coincidence." We consider this remark to be quite misleading: indeed, we point out in [2] that the reproduction of the high-$T$ series in [3] is merely a fit of 11 unknown expansion coefficients (for the weights $w_y$ and $w_z$ ) to ensure agreement with the 11 exactly known high-$T$ terms. Notably, no *further* high-$T$ series coefficients are proposed in [3]; however, since this fit turns out to play no further role, it remains true that the conjectured solution does *not* reproduce the exact high-*T* expansion.

We do not find the majority of the issues addressed in the Response to be relevant to our disproof of [3], which also stressed the failure of the conjectured solution to generate the correct low-*T* expansions. In our view, a refusal to accept the conclusions of the rigorous work (cited in [2]) for the applicability of the long-known expansions — at high enough and low enough *T* — to the exact solution for the thermodynamic limit, constitutes a denial of the mathematical basis of statistical mechanics.